\documentclass{aa}  

\usepackage[utf8]{inputenc}
\usepackage{amssymb,amsmath}
\usepackage{graphicx}
\usepackage[dvipsnames]{xcolor}
\usepackage[breaklinks=true]{hyperref}
\hypersetup{colorlinks=true,allcolors=teal}
\usepackage[flushleft]{threeparttable}
\usepackage{footmisc}
\usepackage{subfigure}
\usepackage{multirow}

\newcommand{\be}{\begin{equation}}
\newcommand{\ee}{\end{equation}}
\def\bear#1\ear{\begin{align}#1\end{align}}

\usepackage[T1]{fontenc}
\usepackage{amsmath}	
\DeclareRobustCommand{\VAN}[3]{#2}
\let\VANthebibliography\thebibliography
\def\thebibliography{\DeclareRobustCommand{\VAN}[3]{##3}\VANthebibliography}
\usepackage{graphicx}
\usepackage{txfonts}
\begin{document}

   \title{An emulator-based forecasting on astrophysics and cosmology with 21 cm and density cross-correlations during EoR}


   \author{Barun Maity
          \inst{1}}

   \institute{Max-Planck-Institut f\"ur Astronomie, K\"onigstuhl 17, D-69117 Heidelberg, Germany\\
              \email{maity@mpia.de}}
              

   \date{Received XXX; accepted XXX}

 
\abstract{The 21 cm signal arising from fluctuations in the neutral hydrogen field, and its cross-correlation with other tracers of cosmic density, are promising probes of the high-redshift Universe. In this study, we assess the potential of the 21 cm power spectrum, along with its cross power spectrum with dark matter density and associated bias, to constrain both astrophysics during the reionization era and the underlying cosmology. Our methodology involves emulating these estimators using an Artificial Neural Network (ANN), enabling efficient exploration of the parameter space. Utilizing a photon-conserving semi-numerical reionization model, we construct emulators at a fixed redshift ($z = 7.0$) for $k$-modes relevant to upcoming telescopes such as SKA-Low. We generate $\sim7000$ training samples by varying both cosmological and astrophysical parameters along with initial conditions, achieving high accuracy when compared to true simulation outputs. While forecasting, the model involves five free parameters: three cosmological ($\Omega_m$, $h$, $\sigma_8$) and two astrophysical (ionizing efficiency, $\zeta$, and minimum halo mass, $M_{\mathrm{min}}$). Using a fiducial model at the mid-reionization stage, we create a mock dataset and perform forecasting with the trained emulators. Assuming a 5\% observational uncertainty combined with emulator error, we find that the 21 cm and 21 cm-density cross power spectra can constrain the Hubble parameter ($h$) to better than 6\% at a confidence interval of 95\%, with tight constraints on the global neutral fraction ($Q_{\mathrm{HI}}$). The inclusion of bias information further improves constraints on $\sigma_8$ (< 10\% at 95\% confidence). Finally, robustness tests with two alternate ionization states and a variant with higher observational uncertainty show that the ionization fractions are still reliably recovered, even when cosmological constraints weaken.}

               
   \titlerunning{inferring astro-cosmo with 21 cm and density cross-correlation emulator}
   \maketitle
%

\section{Introduction}
\label{sec:intro}
The Epoch of Reionization (EoR) signifies the last major phase transition in the cosmic history of our Universe, when it evolves from a mostly neutral to a mostly ionized state \citep[for reviews, see][]{2001PhR...349..125B,2009CSci...97..841C,2018PhR...780....1D,2022arXiv220802260G,2022GReGr..54..102C}. The fluctuation in the neutral hydrogen field during EoR can be potentially traced by the redshifted 21 cm signal, which arises due to the spin flip transition of the neutral hydrogen atoms at the ground state. The signal carries useful information on cosmological and astrophysical properties of this high redshift epoch. This can provide comprehensive answers to the questions about the ionization and thermal state of the high redshift intergalactic medium (IGM), nature of the first ionizing sources, and the timeline of reionization epoch. It can also inform us about cosmic expansion and structure evolution \citep{2012RPPh...75h6901P}.

While the radio interferometers are gradually improving their sensitivity to detect the 21 cm signal, these still face significant challenges in terms of foreground contamination and instrument characterization, allowing only upper limits on the detection. These efforts include independent groups focussing on different telescopes such as the Low Frequency Array \citep[LOFAR;][]{2019MNRAS.488.4271G,2020MNRAS.493.1662M,2025arXiv250305576M}, the Murchison Widefield Array \citep[MWA;][]{2019ApJ...884....1B,2020MNRAS.493.4711T,2025arXiv250509097N}, the Giant Metrewave Radio Telescope \citep[GMRT;][]{2013MNRAS.433..639P}, and the Hydrogen Epoch of Reionization Array \citep[HERA phase I;][]{hera2023} at redshift range $z\sim6-10$. A few projects, such as Owens Valley Long Wavelength Array \citep[OVRO-LWA][]{2019AJ....158...84E}  and New Extension in Nançay Upgrading LOFAR \citep[NenuFAR][]{2021sf2a.conf..211M,2024A&A...681A..62M}, also aim for higher redshifts covering cosmic dawn. The limits have already been exploited to constrain some of the extreme reionization models \citep{ghara2025}. However, with upcoming facilities like SKA-Low (AA* and AA4 configuration), we expect to detect the signal with percentage-level uncertainties.

As 21 cm signal is supposed to be mainly driven by astrophysical processes, most of the EoR studies with 21 cm signal as a probe mainly focus on constraining uncertain astrophysical parameters, keeping the underlying cosmology fixed. Nonetheless, 21 cm signal can also play a crucial role in probing cosmology in combination with other probes like CMB \citep{Mcquinn2006, Liu2016}. However, the studies aiming for cosmological forecasts with 21 cm signal require efficient models to explore astrophysics and cosmology simultaneously. To this end, analytical halo model of reionization \citep{Schneider2023} and machine learning based techniques \citep{2017ApJ...848...23K, Hasan2020} have recently been exploited, highlighting the prospects of constraining cosmology and astrophysics with 21 cm. Specifically, the approach of creating emulators of observables or likelihoods has been reasonably successful in inferring astrophysical parameters from 21 cm power spectra and EoR observables \citep{2017MNRAS.468.3869S,schmidt_pritchard2018,Breitman2024,2024MNRAS.527.9977S,2023MNRAS.526.3920M,2024arXiv240703523C,2024JCAP...03..027C}. 

Parallelly, the windows are now getting open for synergetic studies with 21 cm and other high redshift probes. For example, the distribution of Ly-$\alpha$ emitting high redshift galaxies can be used as a biased tracer of large scale density structure of the universe and is useful for cross-correlation with 21 cm \citep{2016MNRAS.457..666V,Laplante2023,Moriwaki2024}. There exist other tracers of density, such as intensity maps \citep[as a review, see, ][]{2017arXiv170909066K} of CO \citep{2011ApJ...741...70L}, CII \citep{2012ApJ...745...49G}, H$\alpha$ \citep{2021MNRAS.506.1573H}, OIII \citep{2018MNRAS.481L..84M};  which can also be utilized for cross-correlation studies. Hence, the cross-correlation between 21 cm and cosmic matter density, an estimator independent of any specific tracers, can provide us with a potential probe of the high redshift universe \citep{Xu2019}. Furthermore, the cross power spectra is a superior probe in terms of signal-to-noise ratio due to uncorrelated systematics and can complement the pure 21 cm auto power spectra signal. With the availability of current and upcoming telescopes like James Webb Space Telescope (JWST), Nancy Grace Roman Space Telescope (NGRST), Extremely Large Telescope (ELT) etc, the cross-correlation prospects have been shown to be promising in gleaning astrophysical signal during EoR \citep{2025arXiv250220447G}. In principle, the cross-correlation information can also be utilized to infer cosmology, which has already been explored in low redshift studies \citep{2024MNRAS.529.4803B, 2025arXiv250418625A}.

In this study, we aim to check the prospects of 21 cm and its cross-correlation information with matter density in constraining both astrophysics and cosmology during reionization epoch.  Unlike other semi-numerical approach based on the excursion set algorithm, we utilize a more realistic prescription provided by \textbf{S}emi Numerical \textbf{C}ode for \textbf{R}e\textbf{I}onization with \textbf{P}ho\textbf{T}on Conservation (\texttt{SCRIPT}) to generate the neutral hydrogen fluctuation field. We consider only a single redshift ($z=7.0$) for creating the emulators and pursuing parameter exploration, which gives us a starting point as a proof of concept.  However, this can be extended to multiple redshifts, exploring the full power of 21 cm observables in future studies.

The paper is organized as follows: In section \ref{sec:theory}, we describe the reionization model and define the observables explored. Next, we discuss building the emulators of those observables, highlighting the performance of our emulators against the true values in section \ref{sec:emulator}. Once the emulator is trained, we describe the mock generation procedure in section \ref{sec:gen_mock} and parameter exploration in section \ref{sec:param_exp}. Then, we discuss our main results in section \ref{sec:results}. Finally, we summarize the paper in section \ref{sec:conc}.

\section{Reionization model and observables/estimators}
\label{sec:theory}
The reionization models implemented in \texttt{SCRIPT} were originally introduced by \citet{2018MNRAS.481.3821C} and have since been exploited with various observables \citep{2022MNRAS.511.2239M, 2022MNRAS.515..617M}. In this work, we adopt the simplest version of this framework—a two-parameter reionization model previously used for 21 cm forecasting \citep{2023MNRAS.521.4140M}. For completeness, we briefly summarize the methodology here.

\texttt{SCRIPT} simulates the ionization state of the Universe within a cosmologically representative volume, enabling the computation of large-scale ionization fluctuation power spectra that are converged with respect to the resolution of the simulation box \citep{2018MNRAS.481.3821C}. To initialize the model, we provide the density field and the spatial distribution of collapsed halos capable of emitting ionizing radiation. Focusing on large-scale IGM features, we use the second-order Lagrangian perturbation theory (2LPT) to generate the density field, rather than relying on computationally expensive full N-body simulations. Specifically, we used the implementation by \citet{2011MNRAS.415.2101H}\footnote{\url{https://www-n.oca.eu/ohahn/MUSIC/}}. This model also allows us to vary different cosmological parameters ($\Omega_m, h, \sigma_8$, $n_s$, $w_0$ in this case) as well as the initial seed for generating the fluctuating fields. The parameters have the standard meanings i.e., $\Omega_m$: dark matter density, $h$: bubble parameter, $\sigma_8$: quantifies the amplitude of primordial matter fluctuations, $n_s$: tilt of the primordial power spectra, and $w_0$: dark energy equation of state. We fix the baryonic density parameter ($\Omega_b=0.0482$) throughout the study. The distribution of halos is computed using a subgrid approach based on the conditional ellipsoidal collapse mass function \citep{2002MNRAS.329...61S}. Although this approach has been proven to be extremely successful for standard $\Lambda$CDM models, one essential assumption for our study is that the prescription remains the same for the range of cosmology models considered here.  Simulations are conducted within a comoving box of size $256~h^{-1}\mathrm{cMpc}$, which has been shown to be sufficient for the observables considered here, as demonstrated in recent literature \citep{2014MNRAS.439..725I, 2020MNRAS.495.2354K}.
The spatial resolution is set to $\Delta x = 2~h^{-1}\mathrm{cMpc}$, adequate for capturing the scales accessible to SKA-Low.

As mentioned earlier, we use a basic reionization model containing two free parameters, which are needed to get the ionization topology. The model adopts the photon-conserving algorithm to construct the reionization topology within the simulation box. Specifically, the ionization field relies on the ionization efficiency parameter ($\zeta$), which estimates the available ionizing photons per hydrogen atom and minimum threshold halo mass ($M_{\mathrm{min}}$) required to get the fraction of mass collapsed inside a halo. We restrict ourselves to this simple two-parameter setup as we aim to pursue a prospective forecasting study with 21 cm and its cross-correlations with matter density, while simultaneously varying astrophysical and cosmological parameters. This basic setup helps us to gain the required efficiency by minimizing the parameter space. However, the study can be expanded with more physical models of reionization, including recombination and radiative feedback effects \citep{2022MNRAS.511.2239M} in a future project.  

In general, any model of reionization produces the ionized hydrogen fraction $x_{\mathrm{HII}, i}$ in grid cells (represented by the index $i$) inside a simulation volume. The differential brightness temperature (assuming spin temperature is very much larger than CMB temperature) is then given by \citep{1997ApJ...475..429M,2003ApJ...596....1C}
\be
\label{eq:delta_Tb}
\delta T_{b, i} \approx 27~\mathrm{mK}  \left(1 - x_{\mathrm{HII}, i}\right) \Delta_i \left(\frac{1+z}{10}\frac{0.15}{\Omega_{m}h^2}\right)^{1/2} \left(\frac{\Omega_{b}h^2}{0.023}\right),
\ee
where $\Delta_i \equiv \rho_{m, i} / \bar{\rho}_m$ is the ratio of the matter density $\rho_{m,i}$ in the grid cell and the mean matter density $\bar{\rho}_m$.

Given these quantities, the 21 cm power spectrum can be computed as 
\be
P_{21}(k) = \langle\hat\delta_{21}(k)\hat\delta_{21}^*(k)\rangle
\ee
where $\hat\delta_{21}(k)$ is the Fourier transform of the mean-subtracted normalized fluctuation field, $\delta T_{b,i}/\langle \delta T_{b, i} \rangle-1$.

Similarly, the cross power spectra between 21cm field and the matter density field are given by 
\be
P_{21\times\delta}=\langle\hat\delta_{21}(k)\hat\delta_m^*(k)\rangle
 \ee
where $\hat\delta_m(k)$ is the Fourier transform of matter density contrast ($\Delta_i-1$). It is worth highlighting that 21 cm-density cross power spectrum can not be observed directly by any tracers, but these can be derived by observing galaxy distribution and estimating galaxy bias with respect to the background dark matter distribution. {With the assumption of linear galaxy bias, galaxy density is essentially proportional to the matter density \citep{Laplante2023}. The linear bias is expected to be a reasonable approximation for the large-scale modes, which are of interest in this study. Hence, 21 cm along with high redshift galaxy surveys, can be utilized as a direct probe of the cross-correlation. For simplification, we use the term observables even for the indirect estimators unless otherwise specified. 
Throughout the paper, we will work with dimensionless power spectra which are given by  
\begin{equation}
\label{eq:Delta_21}
   \Delta_{X}^2(k) = \frac{k^3 P_{X}(k)}{2 \pi^2},
\end{equation}
where $P_X$ corresponds to the different power spectra ($P_{21}$, $P_{21\times\delta}$) as defined earlier. The 21 cm field and the density field are supposed to be highly anti-correlated at large scales, providing negative cross power. This is expected due to the efficient ionization of high density regions, forming the ionizing sources, and similarly, less efficient ionization at low density regions. Hence, we use the amplitude of cross power spectra ($\vert \Delta_{21\times\delta}^2\vert$) as the probe in this study \citep[see also, ][]{Moriwaki2024}. We can further define the bias of the 21cm-density cross power spectrum with respect to matter power spectrum ($P_{\delta\delta}$) as 
\be
\label{eq:bias}
b_{21\times\delta}^2 (k) = \left\vert\frac{P_{21\times\delta}}{P_{\delta\delta}}\right\vert
\ee
This is relatively easy to estimate due to uncorrelated systematics in the cross spectra than the bias of 21 cm auto power spectra. Hence, we choose this probe instead of the bias of the auto power spectra. We also quantify the state of the IGM by globally averaged neutral fraction, $Q_{\mathrm{HI}}=\langle (1-x_{\mathrm{HII,i}})\Delta_i\rangle $. As discussed earlier, we aim to check the prospects of these observables in constraining astrophysical parameters relevant for EoR as well as inferring the underlying cosmological model. This further demands efficient ways for computation, which is discussed in the next section.


\section{Emulating the observables/estimators}
\label{sec:emulator}
\begin{figure*}
    \centering
    \includegraphics[width=0.9\textwidth]{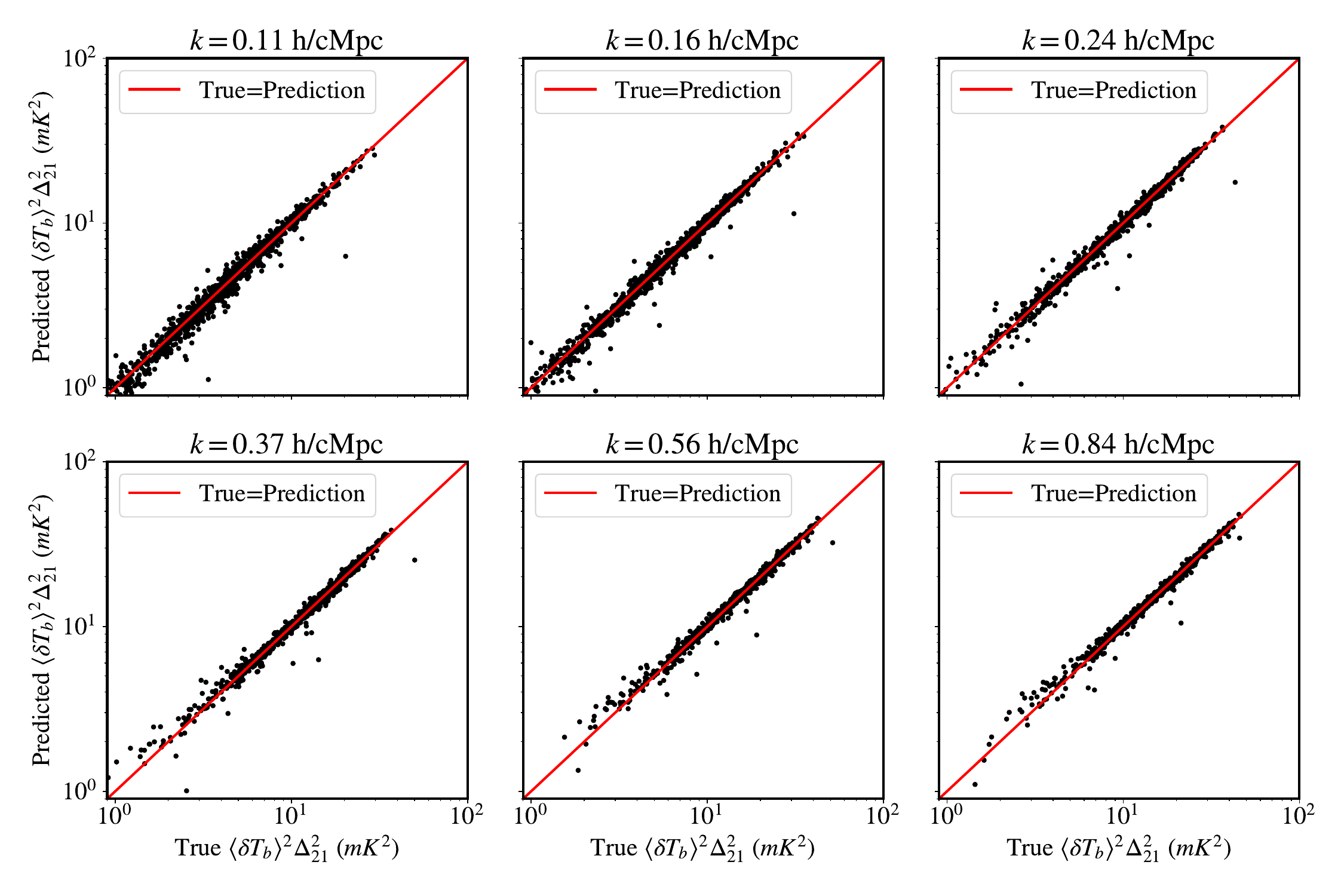}
    \caption{Comparison of true 21 cm power spectrum and corresponding predicted estimates using ANN at different $k$ bins used in this work. The black points correspond to test dataset while the red line signifies True=Prediction. The corresponding $R^2$ value is 0.98.}
    \label{fig:comp_21_pow}
\end{figure*}

\begin{figure*}
    \centering
    \includegraphics[width=0.9\textwidth]{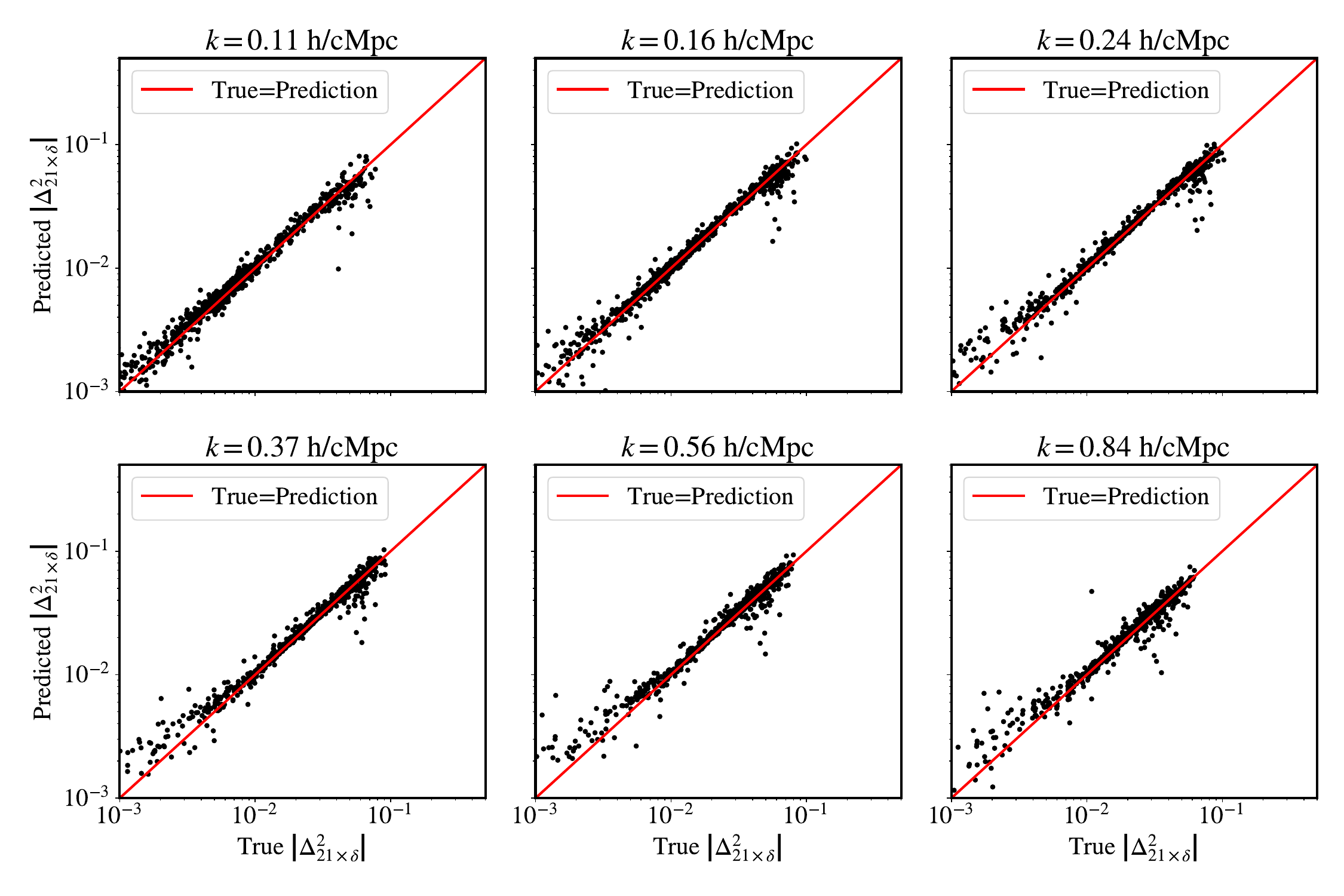}
    \caption{Comparison of true cross power amplitude between 21 cm and $\delta_m$ field with the corresponding predicted estimates using ANN at different $k$ bins used in this work. Other descriptions are similar to Figure \ref{fig:comp_21_pow}. This corresponds to an $R^2$ value of 0.99.}
    \label{fig:comp_21_mat_cross}
\end{figure*}
\begin{figure*}
    \centering
    \includegraphics[width=0.9\textwidth]{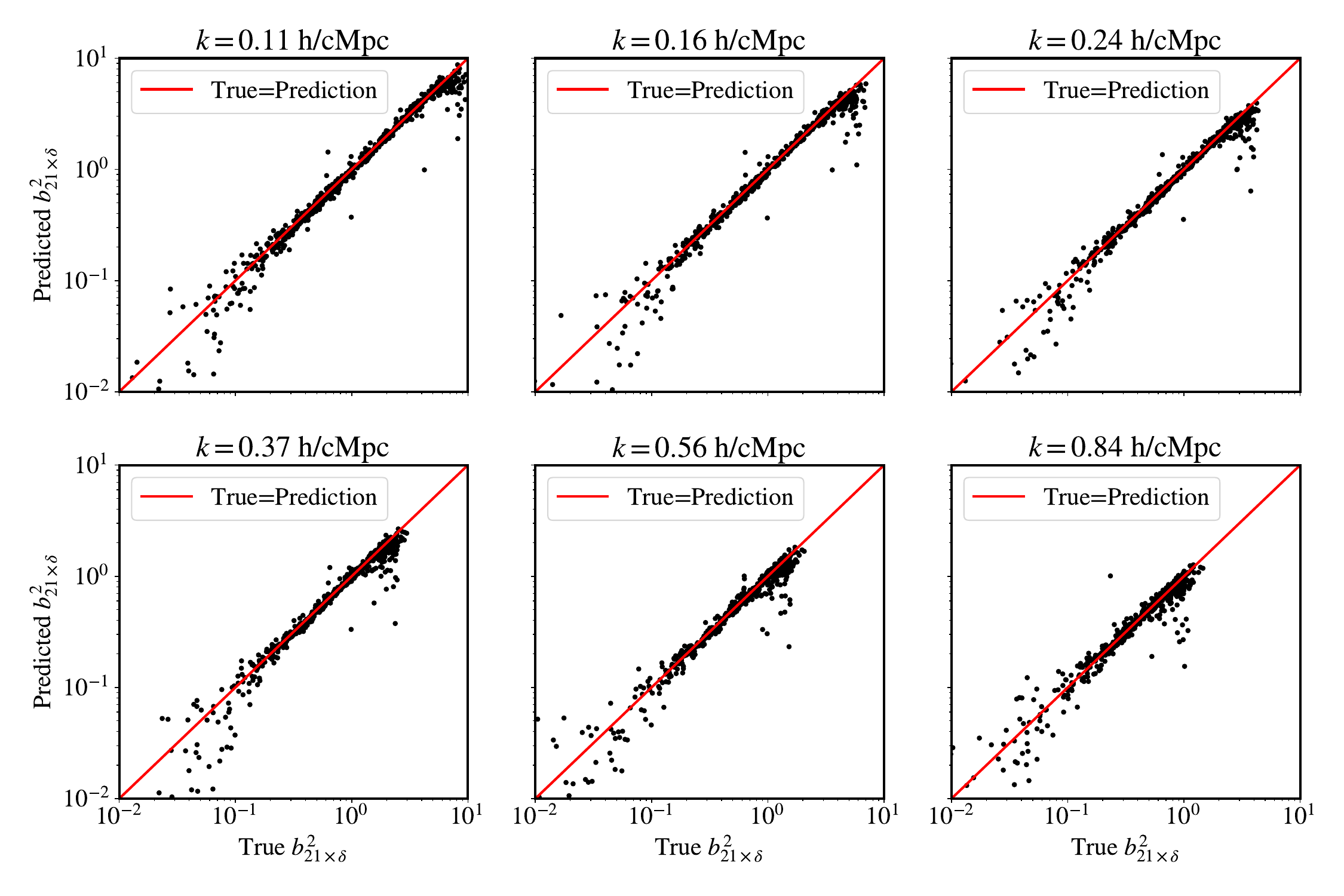}
    \caption{Comparison of true 21 cm bias and the corresponding predicted estimates using ANN at different $k$ bins used in this work.  Other descriptions are similar to Figure \ref{fig:comp_21_pow}. The corresponding $R^2$ value is 0.92.}
    \label{fig:comp_21_bias}
\end{figure*}

\begin{figure*}
    \centering
    \includegraphics[width=0.9\textwidth]{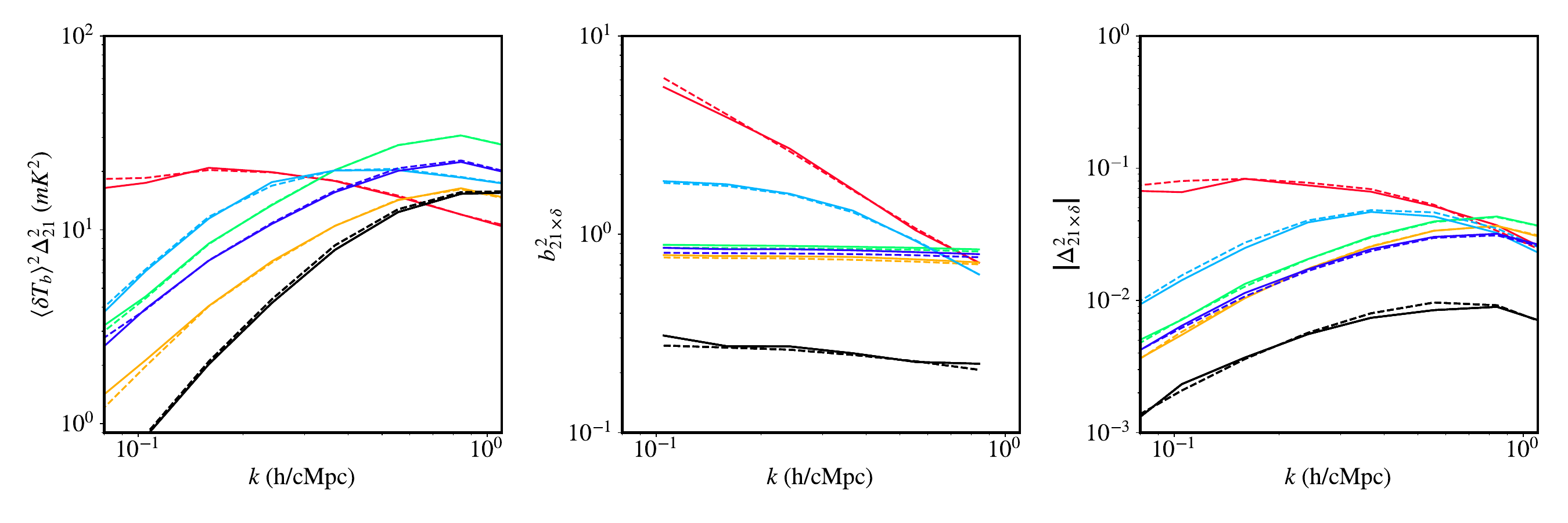}
    \caption{Plots of 21 cm power spectra, 21 cm-density cross power, and its bias for a few random models from the test set. The solid lines are the true models, while the dashed lines are the corresponding predictions.}
    \label{fig:comp_samples}
\end{figure*}
\begin{figure*}
    \centering
    \includegraphics[width=0.9\textwidth]{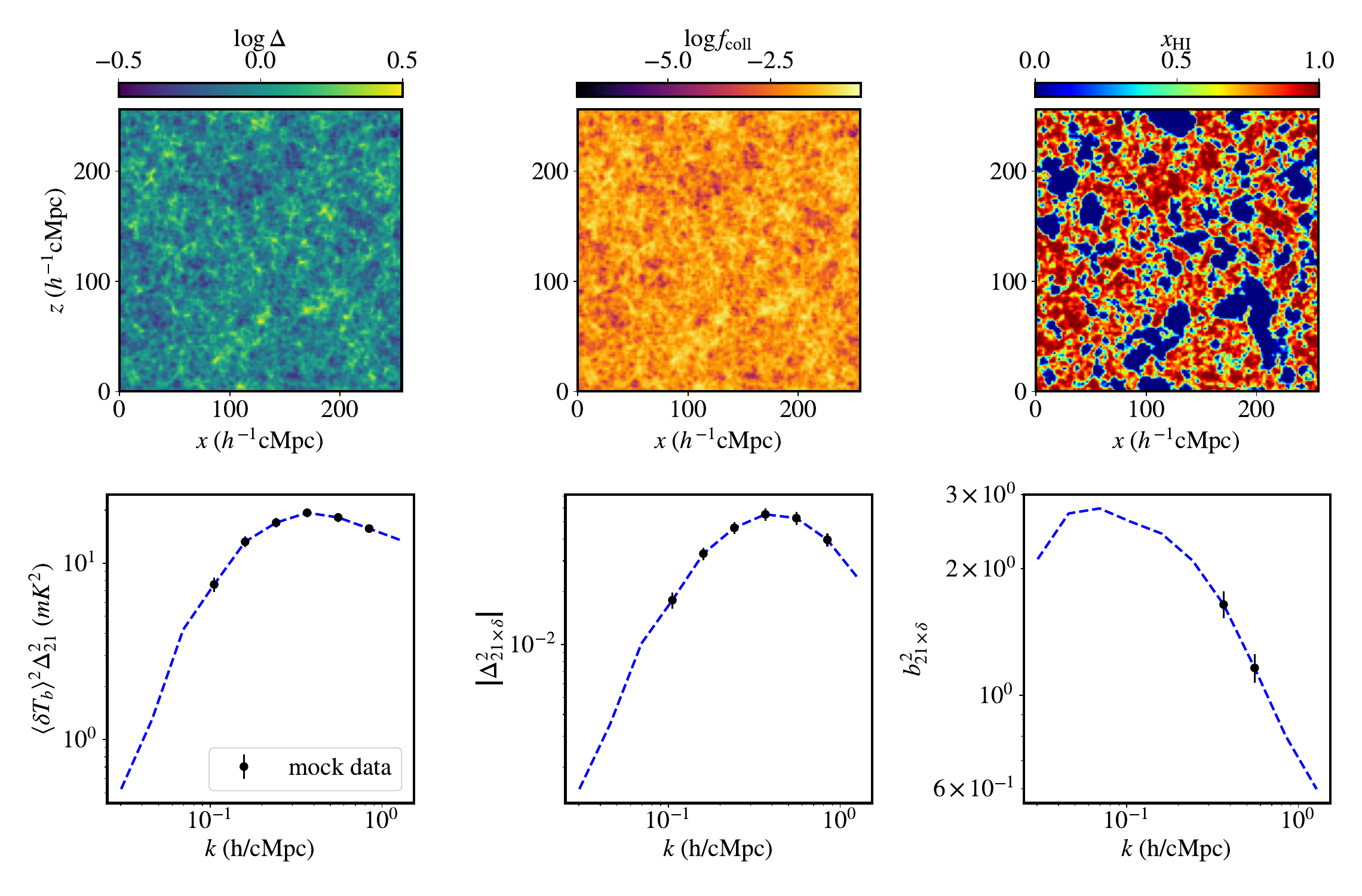}
    \caption{Top panel: Snapshots of Density ($\Delta$), collapsed fraction ($f_{\mathrm{coll}}$), and neutral fraction field ($x_{\mathrm{HI}}$), gradually from left to right, for the fiducial model utilized to generate the mock dataset as described in section \ref{sec:gen_mock}. Bottom panel: The 21 cm power spectra ($\langle\delta T_b\rangle^2\Delta_{21}^2$), 21 cm and density cross power spectra ($\vert \Delta_{21\times\delta}^2\vert$), and the bias of cross spectra ($b^2_{21\times\delta}$) for the corresponding model, gradually from left to right.}
    \label{fig:mock_model}
\end{figure*}
\begin{figure}
    \centering
    \includegraphics[width=0.9\columnwidth]{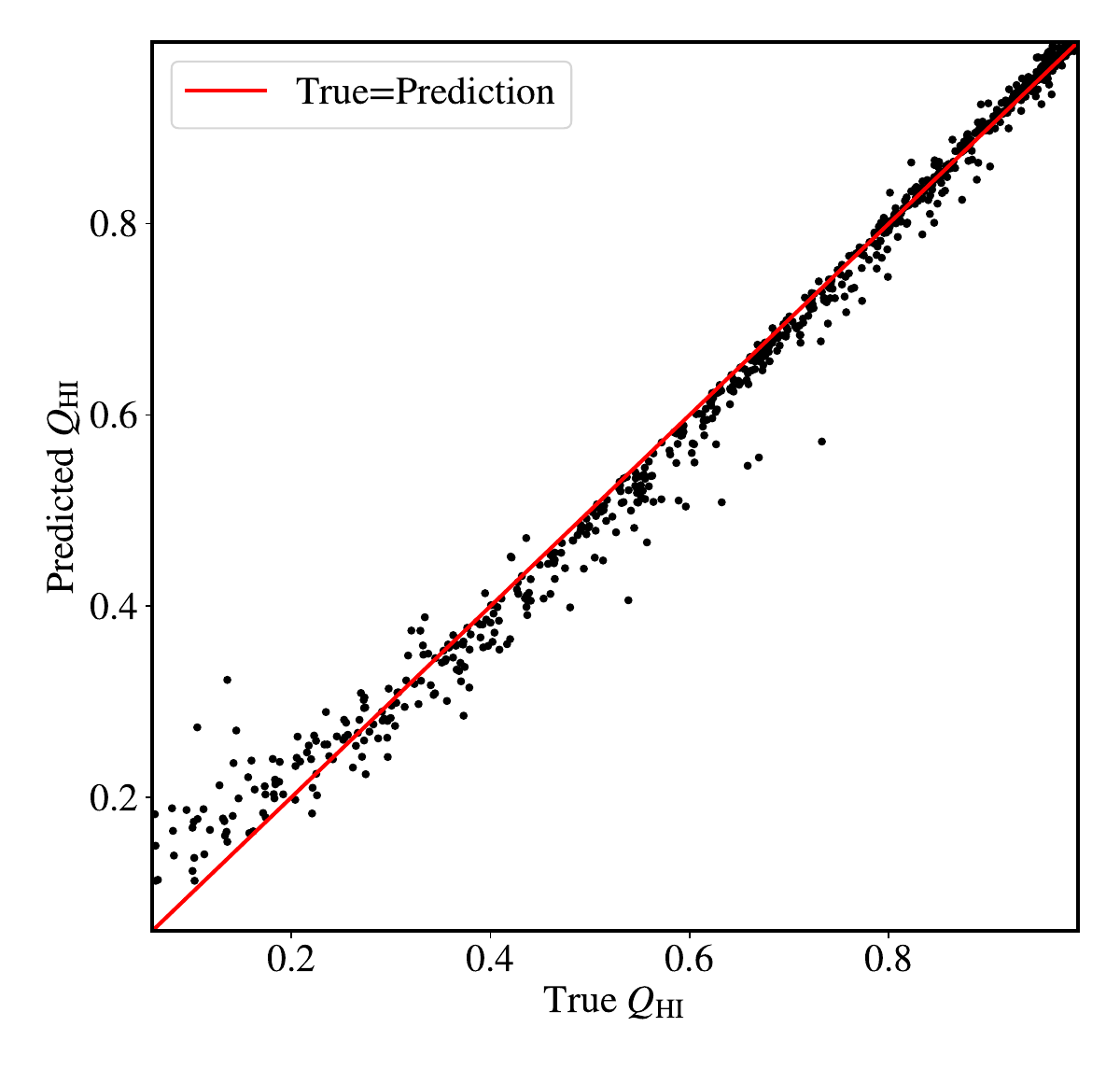}
    \caption{Comparison of True neutral fraction ($Q_{\mathrm{HI}}$) and corresponding Predicted estimates using Gaussian Process Regression.}
    \label{fig:comp_Q}
\end{figure}

In order to pursue parameter space exploration, we construct an emulator which can predict the observables given a set of parameters. In this study, we have two astrophysical parameter ($\zeta, M_{\mathrm{min}}$) defining reionization model and five cosmological parameters ($\Omega_m, h, \sigma_8$, $n_s$, $w_0$) initially. However, we fix $n_s$ and $w_0$ at standard values while pursuing parameter exploration as these are not expected to substantially affect the scales and the redshifts considered here. Hence, we go ahead with rest of the three cosmological parameters, acquiring efficiency. Now, the idea is to predict the observable values given the free parameters as inputs. To this end, we utilize supervised machine learning technique, specifically, an Artificial Neural Network (ANN) to train the emulator.
\subsection{Artificial Neural Network (ANN) in Brief}
 An ANN is composed of an input layer, one or more hidden layers, and an output layer. The input layer receives raw data, while the hidden layers perform complex computations to extract meaningful features. The output layer then provides the final prediction. Each connection between neurons has an associated weight and bias, which are adjusted during the training to minimize errors. Further, in order to allow the network to learn complex patterns, non-linearities are introduced by activation functions, such as ReLU and Sigmoid. The learning process is guided by a loss function, which measures the error, and an optimization algorithm, such as gradient descent or Adam, which updates the weights through backpropagation \citep[for details, see][]{2007JEI....16d9901B,2020MNRAS.491.4031C}. A portion of the training datasets is used for validation purposes during the development of the emulator, and this process progresses in an iterative manner. Once the network is fully trained, it is tested on a different set of data, providing a robustness check on the predictions. In general, ANNs can face two key challenges during training, i.e., underfitting and overfitting. Underfitting occurs when the model is too simple to capture the patterns in the data, leading to poor performance on both the training and validation sets. This often happens when the network has too few layers or neurons, or when the training time is insufficient. On the other side, overfitting happens when the model learns the noise and specific details of the training data instead of generalizing to new data. This results in excellent performance on the training set but poor accuracy on unseen validation or test data. Overfitting is common when the network is too complex or trained for too many epochs without proper regularization. To check the performance of the network, we use $R^2$ metric score, which is defined as 
\be
R^2 = 1 - \frac{\sum (y_{\mathrm{true}}-y_{\mathrm{predict}})^2}{\sum (y_{\mathrm{true}}-\langle y_{\mathrm{true}}\rangle)^2}
\ee
where $y_{\mathrm{true}}$ is the true value of the observables from simulation, $\langle y_{\mathrm{true}}\rangle$ is the average from the test set and $y_{\mathrm{predict}}$ is the corresponding prediction from the network. This essentially provides an assessment for goodness of fit, where metric value can vary from 0 to 1.  As the $R^2$ value gets closer to 1, the prediction capability of the emulator gets better. 
\subsection{Training Procedure}
To generate the training dataset, we vary the different parameters within a reasonable prior ranges i.e. $\Omega_m:[0.2,0.4]$, $h:[0.6,0.8]$, $\sigma_8:[0.7,0.9]$, $n_s:[0.9,1.0]$, $w_0:[-2,0]$, $\zeta:[1,40]$ and $\log M_{\min}:[7,12]$. The baryonic density parameter, $\Omega_b$ is fixed at 0.0482, obeying the findings from CMB spectra \citep{planck_2020}. We store the power spectra in 10 different bins between $k\simeq0.05 ~\mathrm{h/cMpc}$ to $1.084~\mathrm{h/cMpc}$. However, we take only 6 bins in the range $k\simeq0.11 ~\mathrm{h/cMpc}$ to $0.84~\mathrm{h/cMpc}$ for further analysis, which are expected to be probed by upcoming instruments like SKA-Low. The goal here is to predict the corresponding amplitude in those bins given a set of parameter values. Given that motivation, we generate a total of 6750 samples for each type of observables (21 cm power spectra, 21cm-density cross power spectra amplitude and cross bias) by randomly varying these parameters. Among these samples, 500 correspond to different realizations of the initial seed. This further takes into account the cosmic variance uncertainties during the training. We utilize publicly available Scikit-learn and TensorFlow packages in Python to implement the network.  We split the sample in training and testing sets with a ratio of 80 to 20. Our assumed network architecture is summarized in Table \ref{tab:arch}.  We use ReLU activation between the layers and Adam optimizer in this setup. An architecture with 10 hidden layers along with one input and one output layer, performs well to serve the purpose of the study. 
\begin{table}
\centering
\caption{ANN architecture for training the 21 cm power spectra and cross bias}
\renewcommand{\arraystretch}{1.2}
\setlength{\tabcolsep}{10pt}

\begin{threeparttable}
\begin{tabular}{cc}
\hline
Layers & Description \\
\hline
Input & free parameters\\

Dense 1 & (512 neurons, activation='relu')\\

Dense 2 & (1024 neurons, activation='relu')\\

Dense 3 & (1024 neurons, activation='relu')\\

Dense 4 & (512 neurons, activation='relu')\\
   
Dense 5 & (512 neurons, activation='relu')\\
   
Dense 6 & (256 neurons, activation='relu')\\
    
Dense 7 & (128 neurons, activation='relu')\\
    
Dense 8 & (64 neurons, activation='relu')\\
   
Dense 9 & (32 neurons, activation='relu')\\

Dense 10 & (16 neurons, activation='relu')\\
output & (values at $k$ bins, activation='linear')

\end{tabular}
\end{threeparttable}
\label{tab:arch}
\end{table}

In Figure \ref{fig:comp_21_pow}, we show the comparison between true 21 cm power amplitude against the prediction from the trained network for the test samples. Each panel shows the six different bins considered in this study. It is visually clear that the true and predicted values are well correlated with each other, signifying a good accuracy of the prediction. The overall $R^2$ metric score for the test set is 0.98, which also quantifies a well trained model with high predictive power. Similarly, we show the true vs prediction plot for 21 cm-density cross power spectrum in Figure \ref{fig:comp_21_mat_cross} and for cross bias in Figure \ref{fig:comp_21_bias}. The corresponding $R^2$ metric values are 0.99 and 0.92, respectively, which again provides a significantly accurate prediction. For cross power spectra, we emulate the quantity $Q_{\mathrm{HI}}\vert \Delta_{21\times\delta}^2\vert$ at first and then divide by $Q_{\mathrm{HI}}$, where the IGM is not fully ionized. This helps us to avoid any possible divergence due to fully ionized IGM in the training set. The scatter at lower amplitudes arises mainly due to the fact that these correspond to highly ionized states of the IGM and hence there is very little amount of leftover correlation information between 21 cm and the density distribution. In Figure \ref{fig:comp_samples}, we further show the comparison plots of the true model and emulator prediction for the different observables as functions of $k$ modes using six random sets of parameter samples in our test suite. The true (in \textit{solid}) and the predicted (in \textit{dashed}) cases match reasonably well for the different models. This gives us confidence on the emulator's performance over a wide range of models. In Figure \ref{fig:mock_model}, we give an example of a fiducial model which has been discussed in section \ref{sec:gen_mock}

We also utilize the datasets to predict the global neutral fraction ($Q_{\mathrm{HI}}$), providing the same set of free parameters. Instead of a complex network, a simpler technique using Gaussian Process Regression (GPR) is sufficient to give a reasonably accurate prediction in this case, corresponding to $R^2$ metric score of 0.98.  In Figure \ref{fig:comp_Q}, we show the comparison between the true global neutral fraction and the corresponding predictions from GPR. These are nicely correlated with each other along the equality line with an average scatter uncertainty $<5\%$.  

\begin{figure*}
    \sidecaption
    \centering
    \includegraphics[width=0.9\textwidth]{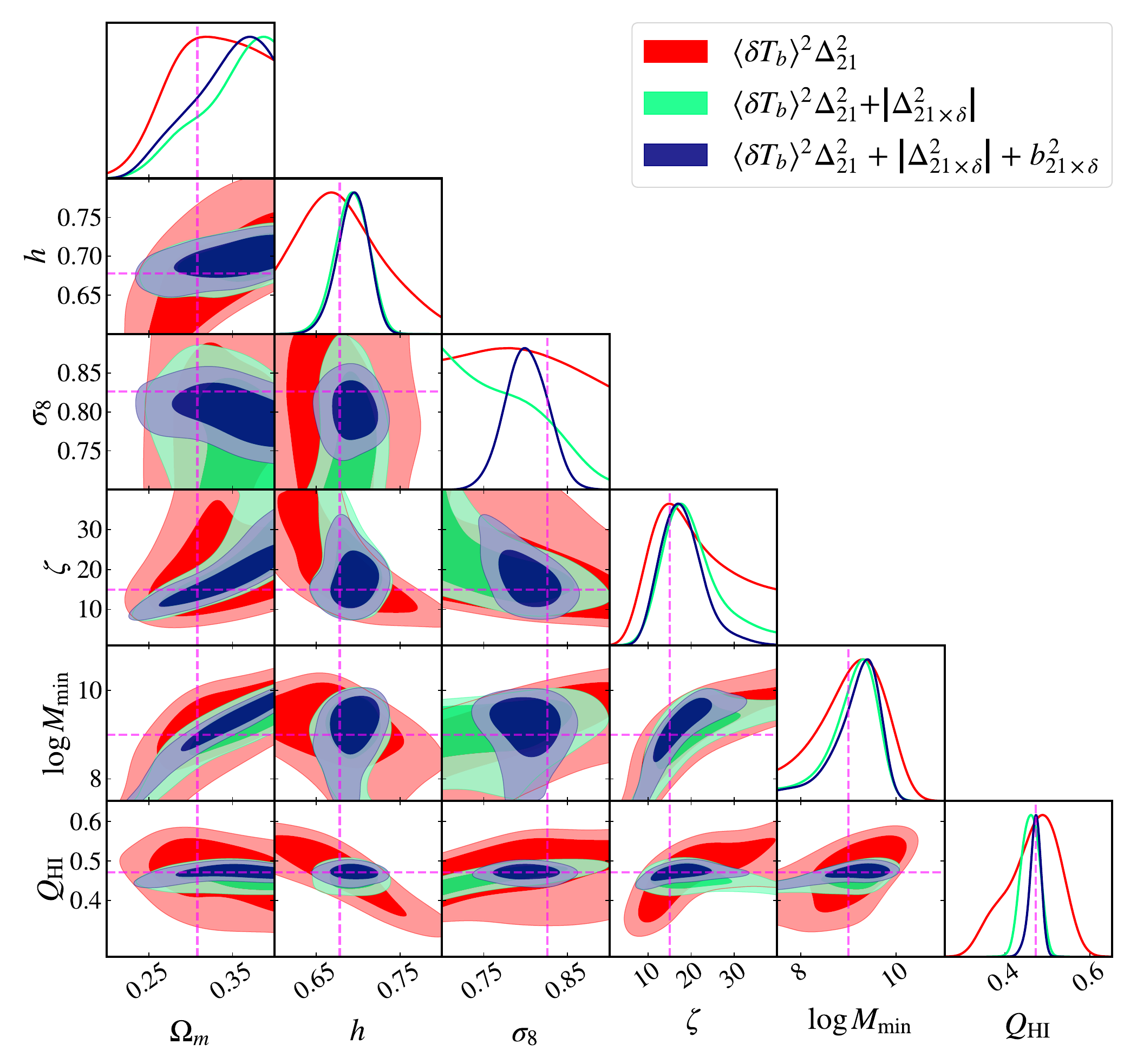}
    \caption{Comparison of posterior distributions using different combinations of observables i.e. only 21 cm power spectra (red), 21 cm power spectra + 21cm-density cross power spectra (green), and adding bias of cross spectra (blue). The diagonal panels show the 1D posterior probability distribution, and the off diagonal panels show the joint 2D posteriors. The contours represent the 68\% and 95\% confidence intervals. The dashed line represents the input parameter values used to generate the mock dataset. The observational uncertainties are assumed to be 5\% of the observable amplitudes in this case.}
    \label{fig:comp_posterior_fiducial}
\end{figure*}
\begin{figure*}
    \sidecaption
    \centering
    \includegraphics[width=0.9\textwidth]{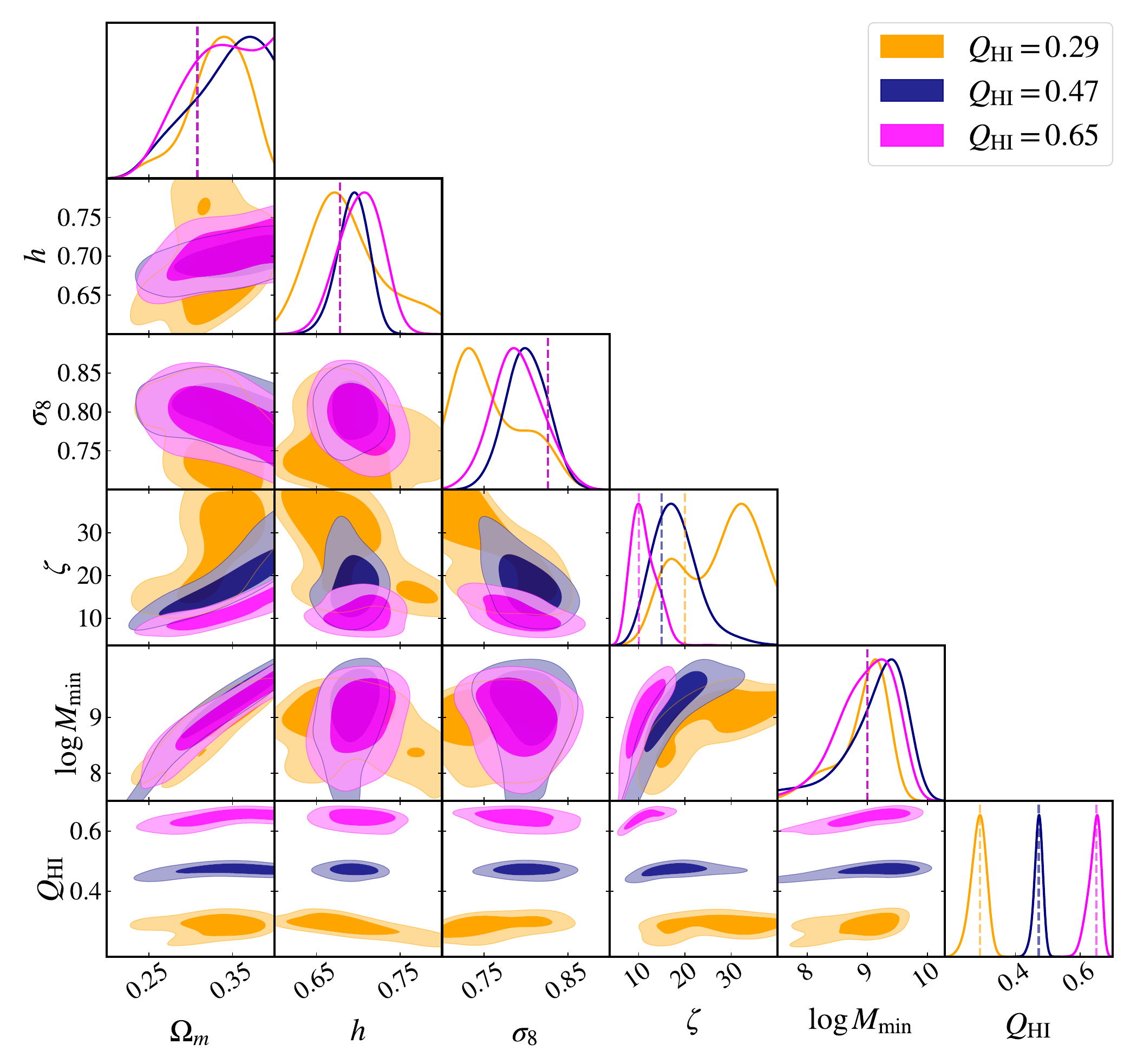}
    \caption{Comparison of posterior distributions for mocks corresponding to different ionization states. Along with the fiducial case ($Q_{\mathrm{HI}}=0.47$, in blue), we show two more cases for lower ($Q_{\mathrm{HI}}=0.29$, in orange)  and higher neutral fraction ($Q_{\mathrm{HI}}=0.65$, in magenta). The observational uncertainties are assumed to be 5\% of the amplitudes as in Figure \ref{fig:comp_posterior_fiducial}.}
    \label{fig:comp_posterior_QHI}
\end{figure*}
\begin{figure*}
    \sidecaption
    \centering
    \includegraphics[width=0.9\textwidth]{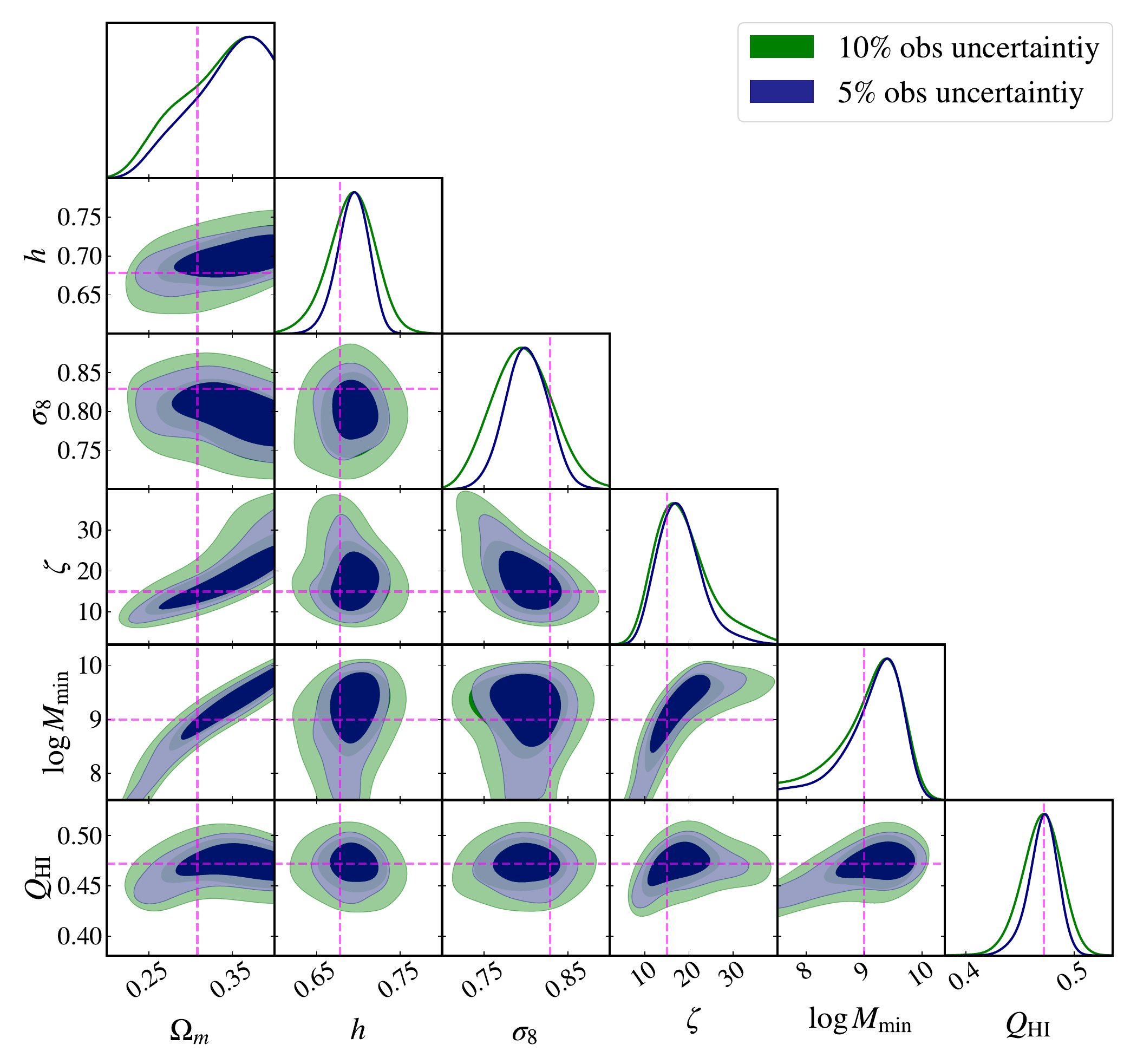}
    \caption{Comparison of posterior distributions for mocks corresponding to different observational uncertainties assumed in the likelihood. The blue contours correspond to the fiducial case with 5\% uncertainties, as also shown in Figure \ref{fig:comp_posterior_fiducial} and \ref{fig:comp_posterior_QHI}; while the green contours correspond to a variant with 10\% observational uncertainties for all the observables used.}
    \label{fig:comp_posterior_uncertain}
\end{figure*}
\section{Generating the mock data} 
\label{sec:gen_mock}
We choose a fiducial set of parameters to generate the mock observables. For exploration studies, we fix $n_s=0.961$ and $w_0=-1$, consistent with the CMB estimates \citep{planck_2020}. The rest of the parameter values corresponding to these fiducial mocks are chosen as $\Omega_m=0.308$, $h=0.678$, $\sigma_8=0.829$, $\zeta=15$, and $\log M_{\mathrm{min}}=9.0$. The values are shifted by a Gaussian random noise with a standard deviation consistent with the expected level of SKA-Low (AA*) thermal noise with 1000 hrs of observations. These shifted values are treated as the mock dataset for our analysis. The total errors on the mock dataset are assumed to have two contributions i.e. training uncertainties ($\sigma_\mathcal{D}^{tr}$) and the overall observational uncertainties ($\sigma_\mathcal{D}^{obs}$). We compute the training uncertainties by quantifying the scatter in True vs Predicted observable distributions. Specifically, we estimate 84\% and 16\% quantiles for (True-Predicted) distributions and then take half the difference between those two quantiles to get $\sigma_\mathcal{D}^{tr}$. For observational uncertainties, we assumed a moderate value, 5\% of the observable amplitude at the corresponding $k$ bins. This is motivated by the expected SNR ($\gtrsim20$) on 21 cm power spectra from SKA-Low AA* observations for 1000 hrs with an optimistic foreground scenario \footnote{\url{https://21cmsense.readthedocs.io/en/latest/tutorials/SKA_forecast.html}}. We also show a case with a more conservative uncertainty, assuming 10\% of the amplitude.  Then the total uncertainties are estimated by adding these  contributions in quadrature as
\be
\sigma_\mathcal{D}^{tot} = \sqrt{(\sigma_\mathcal{D}^{tr})^2 + (\sigma_\mathcal{D}^{obs})^2}
\ee

At the top panels in Figure \ref{fig:mock_model}, we show the different cosmological fields, including matter density ($\Delta$), collapsed fraction ($f_{\mathrm{coll}}$), and neutral fraction ($x_{\mathrm{HI}}$). The fluctuations in these fields are nicely correlated with each other. At the bottom panels, we show the corresponding observables i.e. 21 cm power spectra ($\langle\delta T_b\rangle^2\Delta_{21}^2$), 21cm-density cross power spectra ($\vert\Delta_{21\times\delta}^2\vert$), and bias ($b^2_{21\times\delta}$). The black data points with errorbars are utilized for the parameter space exploration studies, while the blue dashed line represents the underlying input model. We adopt a conservative approach by using only two $k$ bins with the better prediction uncertainties (among all the 6) for the 21 cm bias, as this has a relatively lower prediction accuracy among the three estimators considered in this study. This helps us to minimize the bias coming from poor emulator predictions.

We further study utilizing two variants of the mock dataset using different input models corresponding to different ionization states. These are generated by changing the $\zeta$ values appropriately. To ensure robustness,  we check the results with a smaller ($\zeta=10$) and a larger ($\zeta=20$) value than the fiducial ones as discussed later in section \ref{sec:results}.

\section{Parameter exploration with emulator}
\label{sec:param_exp}
We employed a standard Bayesian framework to explore the parameter space of our model. Our objective was to compute the posterior probability distribution, $\mathcal{P}(\lambda \vert \mathcal{D})$, of the model parameters $\lambda$, conditioned on the observational (mocks in this case) datasets $\mathcal{D}$ introduced in the previous section. According to Bayes’ theorem, the posterior is given by

\begin{equation}
\label{eq:bayes_eq}
\mathcal{P}(\lambda \vert \mathcal{D}) = \frac{\mathcal{L}(\mathcal{D} \vert \lambda), \pi(\lambda)}{\mathcal{P}(\mathcal{D})},
\end{equation}

where $\mathcal{L}(\mathcal{D} \vert \lambda)$ denotes the likelihood, $\pi(\lambda)$ represents the prior distribution, and $\mathcal{P}(\mathcal{D})$ is the Bayesian evidence. The evidence serves as a normalization constant and does not influence our parameter inference.

The likelihood function was modeled as a multivariate Gaussian distribution,
\be
\label{eq:chisq_eq}
\mathcal{L}(\mathcal{D} \vert \lambda) 
=\exp \left(-\frac{1}{2} \sum_{\alpha}\left[\frac{\mathcal{D}(k_{\alpha})-\mathcal{M}(k_{\alpha}; \lambda)}{\sigma_\mathcal{D}^{tot}(k_{\alpha})}\right]^2 \right)
\ee
Here $\mathcal{D}$ corresponds to different mock observables/estimators (i.e. $\langle\delta T_b\rangle^2\Delta_{21}^2$, $\vert \Delta_{21\times\delta}^2\vert$ and $b^2_{21\times\delta}$) and $\mathcal{M}$ is the corresponding predictions for parameter set $\lambda$.

To sample the posterior distribution, we utilized the Markov Chain Monte Carlo (MCMC) method, employing the Metropolis–Hastings algorithm \citep{1953JChPh..21.1087M}. The MCMC chains were executed using the publicly available \texttt{cobaya} package \citep{2021JCAP...05..057T}\footnote{\url{https://cobaya.readthedocs.io/en/latest/}}.  
We checked the convergence of the chains following the  Gelman-Rubin $R - 1$ statistic \citep{1992StaSc...7..457G}. The chain was assumed to have converged when the $R - 1$ value was lower than a threshold $0.01$. 
For subsequent analysis, we discarded the initial 30\% of samples from each chain as burn-in and based our inference on the remaining samples.
\begin{table*}
\centering
\caption{Parameter constraints obtained from the MCMC-based analysis for different scenarios using mock dataset.}
\renewcommand{\arraystretch}{0.5}
\setlength{\tabcolsep}{5.5pt}
\small
\begin{threeparttable}
\begin{tabular}{cccccccc}
\hline
\\
Parameters  & Input & \multicolumn{4}{c}{$\langle\delta T_b\rangle^2\Delta_{21}^2 + \vert \Delta_{21\times\delta}^2\vert + b^2_{21\times\delta}$} & $\langle\delta T_b\rangle^2\Delta_{21}^2 + \vert \Delta_{21\times\delta}^2\vert$ & $\langle\delta T_b\rangle^2\Delta_{21}^2$   \\

&  &  & &&&& \\
(95\% limits) &  &  ($Q_{\mathrm{HI}}=0.47$)  &  ($Q_{\mathrm{HI}}=0.47$) & ($Q_{\mathrm{HI}}=0.65$)& ($Q_{\mathrm{HI}}=0.29$)& 
  ($Q_{\mathrm{HI}}=0.47$) &  ($Q_{\mathrm{HI}}=0.47$)  \\ 
 &  &  & &&&& \\
&  & ( 5\% obs. err.) & ( 10\% obs. err.) & ( 5\% obs. err.) & ( 5\% obs. err.) & ( 5\% obs. err.) & ( 5\% obs. err.)\\\\
\hline 
&  &  & &&&& \\
 \hline
&  &  & &&&& \\

  $\Omega_m$ &0.308  &[$>0.26]$   & [$>0.25$]    & [$>0.26$] & [$>0.26$]  & [$>0.27$] & [$>0.25$]\\ \\
  $h$ & 0.678 &0.69 [0.66, 0.73]    & 0.69 [0.63, 0.74]     & 0.70 [0.65, 0.75] & 0.68 [0.61,0.78] & 0.69 [0.65,0.73] & [$<0.77$]\\ \\
    $\sigma_8$ & 0.829 &0.80 [0.75, 0.85]   & 0.80 [0.73, 0.86]     & 0.79 [0.73, 0.85] & [$<0.83$] &[$<0.86$] & -\\ \\
$Q_{\mathrm{HI}}$& - & 0.47 [0.44, 0.50]   & 0.47 [0.43, 0.50]     & 0.64 [0.60, 0.68] & 0.29 [0.23, 0.33] & 0.46 [0.42, 0.50] & 0.46 [0.34, 0.57] \\ \\

\hline \\

\end{tabular}
\end{threeparttable}
\label{tab:param_cons}
\tablefoot{For each case, we show the 95\% confidence limits on the parameters in the brackets. We provide the mean posterior values where the bounds are available from both sides. }
\end{table*}
\section{Results}
\label{sec:results}
In this section, we discuss the findings from our parameter space exploration studies using mock datasets. In Table \ref{tab:param_cons}, we provide the 95\% confidence limits along with mean of the recovered cosmological parameters as well as the ionization state for different scenarios considered in this study.

In Figure \ref{fig:comp_posterior_fiducial}, we show the posterior recoveries of the free parameters from the fiducial mock dataset. The free parameters include both cosmological ($\Omega_m$, $h$, $\sigma_8$) as well as astrophysical ones ($\zeta$, $\log M_{\mathrm{min}}$). We also show the posterior of the globally averaged neutral fraction ($Q_{\mathrm{HI}}$) as a derived parameter. We find that the parameters are not well constrained for the case where we utilize only 21 cm power spectra (shown in red), although it can correctly recover the global neutral fraction with wide uncertainties. The constraints are improved significantly when we include 21cm-density cross power spectra as observables along with 21 cm power spectra (shown in lime green). Specifically, the Hubble parameter ($h$) is constrained within an uncertainty of $<6\%$ at a confidence interval of 95\%. This signifies the potential of 21 cm and synergies with galaxy observables as an independent probe to constrain the expansion rate of the universe, which can further shed light onto the well known Hubble tension \citep{2019ApJ...876...85R, 2019NatAs...3..891V}. Furthermore, the global neutral fraction is now stringently constrained, discarding a significant portion of astrophysical parameter spaces. On top of that, if we include the bias of cross spectra, it further constrains $\sigma_8$ parameter (providing $<10\%$ uncertainty at 95\% confidence). This happens as the combination now has the information on the amplitude of the underlying matter power spectra (see equation \ref{eq:bias}), which is controlled by $\sigma_8$. We also note that the reionization source parameters are also well recovered, and the uncertainties subsequently improve as we include more observables. However, $\Omega_m$ is bounded by only one side due to strong degeneracy with astrophysical parameters.

To check the robustness of the findings, we further pursue parameter space exploration using two more mock datasets with different ionization states. We tune the ionizing efficiency parameter to generate the mocks with a higher ($Q_{\mathrm{HI}}=0.65$) and a lower ($Q_{\mathrm{HI}}=0.29$) neutral fraction than the fiducial one ($Q_{\mathrm{HI}}=0.47$), while all the other input parameters are kept the same as before. In Figure \ref{fig:comp_posterior_QHI}, we show the recovered posteriors of these cases along with the fiducial one. We find that the $h$ and $\sigma_8$ parameters are well constrained even for the higher neutral fraction, recovering the underlying true values within 95\% uncertainties as before. The astrophysical parameters are also consistent with the input values. On the other hand, the constraints on the parameters for lower neutral fraction are not significantly strong, barely constraining $h$ and providing one sided bound on $\sigma_8$ at 95\% uncertainty level. This is not very surprising as the ionized bubbles start to overlap when the universe is highly ionized (lower neutral fraction) which can wipe out the correlation information, resulting in a loss of constraining power.  However, the neutral fraction has still been recovered with significant precision without any strong bias. This also confirms the fact that the 21 cm observables are more sensitive to the ionization state of the universe rather than the underlying cosmological information. 

Lastly, we check the effects of observational uncertainties on the posterior distribution in Figure \ref{fig:comp_posterior_uncertain}. The green contours show the case where we assume the uncertainties to be 10\% of the observable amplitudes, while the other one is same as the fiducial case with 5\% uncertainties. Not surprisingly, the contour widens for larger uncertainties, however, it still manages to correctly recover the Hubble parameter and amplitude of primordial fluctuations with significant confidence.

\section{Summary and conclusions}
\label{sec:conc}

The astrophysics during the Epoch of Reionization is gradually getting explored with the help of multi-wavelength observables. The 21 cm signal is one of the crucial probes which has the potential to detect neutral hydrogen fluctuations at EoR directly. This further contains useful information about the cosmological parameters, although it is hard to infer cosmology from this weak signal, affected by foreground contamination and poorly understood high redshift astrophysical phenomena. To this end, the cross-correlation of 21 cm signal with other tracers of cosmological density can be a complementary probe of astrophysics and cosmology. This is useful to avoid any systematics arising due to spurious correlation and, hence, enhance the signal to noise ratio of detection. In this study, we check the prospects of 21 cm-density cross power and its bias along with 21 cm power spectra in order to probe the astrophysics and cosmology from the EoR. Our approach relies on creating an efficient emulator of the observables and utilizing the emulator for further parameter space exploration. Below, we summarize this work, highlighting the main findings.


\begin{itemize}
    \item We used a realistic semi-numerical reionization model based on a photon-conserving algorithm to study the prospects of 21 cm and related observables to infer cosmology and astrophysics during the EoR. Specifically, we used 21 cm auto power spectra, magnitude of cross power between  21 cm fluctuations and matter density, and the corresponding bias magnitude. As a prospective study, we focused only at a single redshift i.e. $z=7.0$ in this work. While 21 cm auto power spectra can be observed directly by the radio interferometers,  21cm-density cross power spectra and the bias can not be measured directly. 
    However, the cross power and its bias can be in principle be estimated by different tracers, especially via galaxy-21cm cross-correlation. 
    
    We created a total of  $\sim7000$ samples by varying different astrophysical and cosmological parameters to build the emulator for these observables/estimators. The samples were generated with different initial random seeds, which further takes into account for the cosmic variance uncertainties. The emulators were trained to predict the observables at 6 different $k$ bins, given a set of input free parameters (including astrophysical and cosmological ones). The bins were chosen in the range where we can expect the detection of 21 cm signal from the upcoming telescopes like SKA-Low. We compared the emulators against true values and found that the predictions are sufficiently accurate, providing $R^2$ values $>0.9$ for all the cases (0.98 for 21 cm auto power spectra, 0.99 for cross power spectra, 0.92 for bias amplitude). This provided us with the confidence to do efficient parameter space exploration utilizing the emulators.

    \item Next, we generated the mock observables with a fiducial set of parameter values, consistent with \citet{planck_2020} and providing an ionization state close to the middle of reionization process. We found that 21 cm power spectra alone can not constrain the cosmological parameters, while they can recover the correct ionization state. When we included cross power spectra as another observable, Hubble parameter was constrained and adding bias magnitude on top of it further constrained the amplitude of primordial matter fluctuations ($\sigma_8$). Similarly, the constraints on the ionization state were also improved significantly. We further pursued a similar analysis with two more mock datasets corresponding to a higher and lower neutral fraction. The recoveries were degraded for lower neutral fraction due to a possible lack of correlation information. However, the ionization states were still precisely recovered for all the cases.

\end{itemize}
We would like to caution that the exact quantification of constraints is dependent on the emulator uncertainties, which can be improved with larger datasets spanning wider parameter spaces and with more sophisticated training techniques. We also neglect any covariance between the Fourier modes as well as between observables while computing the likelihood. While the mutual covariances would be ideal to include and may probably degrade the uncertainties, the estimates of 21 cm observables are generally provided without the covariances information in the literature \citep{ghara2025}. Similarly, multi-observable inference studies usually neglect covariance information between the observables \citep{2022MNRAS.515..617M,2025PASA...42...49Q}. Hence, we proceed with the diagonal terms, assuming relatively conservative uncertainties and mutually independent observables. Further, one needs to be cautious while interpreting cosmology with real observational data, as the observational estimates (limits till now) are often derived assuming an underlying cosmology, which should be properly quantified and corrected before inference. All these aspects may be important and will be useful to check with a separate study in the future. To this end, the main conclusions of this study utilizing mocks are unlikely to be changed much, providing an insight into the applicability of 21 cm and corresponding synergies as EoR/cosmology probe. 

The detection prospect of cross power spectrum signal between 21 cm and galaxies are very bright, given the ongoing and upcoming major observational facilities like HERA, SKA, ELT, NGRST etc. For example, HERA-NGRST cross-correlation can provide a 14$\sigma$ detection with an assumption 500 square-deg common survey area \citep{Laplante2023}, while the detection can be improved to 55$\sigma$ with SKA-Low AA* \citep{2025arXiv250220447G}. There also exists exciting potential for cross-correlation between intensity maps of metals like CII, CO and 21 cm, where a $\sim7\sigma$ detection is possible with available instruments \citep{2024ApJ...975..222F}.  To this end, our study provides an expectation on astrophysical as well as cosmological inference during reionization from 21 cm, its cross-correlation with dark matter density and corresponding bias. We chose the direct cross-correlation between 21 cm and density to avoid any further astrophysical uncertainties associated with any specific tracers which also helps to build up efficient emulators.  Although the cross-correlation between 21 cm and dark matter density can not be measured directly, its bias can be derived and these can be useful indirect estimators utilizing the future observations.
 
 Currently, we use a simplistic two parameter reionization model in this study while the realistic universe is expected to be much complex. For example, in a more realistic model, one needs additional parameters such as the IGM clumping factor and temperature increment for photoionization heating, to quantify the effect of the inhomogeneous recombination and radiative feedback processes \citep{2022MNRAS.511.2239M}.  As a natural consequence of a more complex model, the number of training samples is expected to be larger to capture the whole parameter space, along with additional degeneracies between the parameters.  Some of the degeneracies can be alleviated by utilizing complementary reionization probes such as UV luminosity function (UVLFs), Ly-$\alpha$ forest fluctuations,  CMB scattering optical depth etc \citep{2021MNRAS.506.2390Q,2022MNRAS.515..617M,2025PASA...42...49Q}.  However, a two parameter vanilla model is often sufficient to provide the typical nature and amplitude of the fluctuations in the 21 cm field and the state of the IGM \citep{2023MNRAS.521.4140M}, which serves the purpose of this proof of concept study. In the future, we would like to explore with a more realistic reionization model, including above mentioned effects if those affect the cosmological inference. Similarly, the prospects for more direct tracers such as cross correlation between 21 cm and Ly-$\alpha$ emitters density (instead of dark matter density), can be explored, avoiding any assumption of linear scale independent galaxy bias. Parallelly, we would like to extend the study with more redshifts, incorporating full information of high redshift 21 cm observations. Eventually, these can be jointly explored with other EoR and cosmic dawn probes to simultaneously constrain astrophysics and cosmology.


\begin{acknowledgements}
     The author thanks Prof. Tirthankar Roy Choudhury for the comments on the draft, which have been helpful to improve the manuscript.
\end{acknowledgements}

-------------------------------------------------------------------
\bibliographystyle{aa}
\bibliography{script_emu}
\end{document}